\documentclass[twocolumn,english,pra]{revtex4}
\usepackage[T1]{fontenc}
\usepackage[latin9]{inputenc}
\usepackage{amsmath}
\usepackage{graphicx}
\usepackage{amssymb}
\usepackage{esint}

\makeatletter
\@ifundefined{textcolor}{}
{%
 \definecolor{BLACK}{gray}{0}
 \definecolor{WHITE}{gray}{1}
 \definecolor{RED}{rgb}{1,0,0}
 \definecolor{GREEN}{rgb}{0,1,0}
 \definecolor{BLUE}{rgb}{0,0,1}
 \definecolor{CYAN}{cmyk}{1,0,0,0}
 \definecolor{MAGENTA}{cmyk}{0,1,0,0}
 \definecolor{YELLOW}{cmyk}{0,0,1,0}
 }

\makeatother

\makeatother

\usepackage{babel}

\makeatother

\usepackage{babel}

\makeatother

\usepackage{babel}

\makeatother

\usepackage{babel}

\makeatother

\usepackage{babel}

\begin{document}

\title{Virial expansion for a strongly correlated Fermi gas with imbalanced
spin populations}

\author{Xia-Ji Liu$^{1}$}

\email{xiajiliu@swin.edu.au}

\affiliation{$^{1}$ARC Centre of Excellence for Quantum-Atom Optics, Centre for
Atom Optics and Ultrafast Spectroscopy, Swinburne University of Technology,
Melbourne 3122, Australia}

\author{Hui Hu$^{1}$}

\email{hhu@swin.edu.au}

\affiliation{$^{1}$ARC Centre of Excellence for Quantum-Atom Optics, Centre for
Atom Optics and Ultrafast Spectroscopy, Swinburne University of Technology,
Melbourne 3122, Australia}

\date{\today{}}
\begin{abstract}
Quantum virial expansion provides an ideal tool to investigate the
high-temperature properties of a strongly correlated Fermi gas. Here,
we construct the virial expansion in the presence of spin population
imbalance. Up to the third order, we calculate the high-temperature
free energy of a unitary Fermi gas as a function of spin imbalance,
with infinitely large, attractive or repulsive interactions. In the
latter repulsive case, we show that there is no itinerant ferromagnetism
when quantum virial expansion is applicable. We therefore estimate
an upper bound for the ferromagnetic transition temperature $T_{c}$.
For a harmonically trapped Fermi gas at unitarity, we find that $(T_{c})_{uppper}<T_{F}$,
where $T_{F}$ is the Fermi temperature at the center of the trap.
Our result for the high-temperature equations of state may confront
future high-precision thermodynamic measurements. 
\end{abstract}


\maketitle

\section{Introduction}

Strongly interacting Fermi gases of ultracold atoms near Feshbach
resonances have offered unique opportunity to study some long-sought
problems in condensed matter physics, astrophysics and particle physics
\cite{rmpBloch,rmpGiorgini}. An particular example that attracts
intense interests is the ground state of a two-component, spin-population
imbalanced Fermi gas of $^{6}$Li or $^{40}$K atoms \cite{rppRadzihovsky,Chevy}.
In the strongly interacting regime, observation of new exotic superfluid
states in the imbalanced systems may be anticipated, such as the spatially
inhomogeneous Fulde-Ferrell-Larkin-Ovchinnikov state \cite{FF,LO,Orso,hldprl2007,Guan,lhdpra2007,lhdpra2008,riceNature}.
However, because of strong interactions, theoretical investigation
of imbalanced Fermi systems turns out to be difficult and restrictive.

In this work, we theoretically investigate the high-temperature equation
of state of a strongly interacting, imbalanced Fermi gas, by generalizing
a quantum virial expansion method to imbalanced systems \cite{Ho,plbHorowitzQVE,lhdprl2009,lhdpra2010,hldprl2010,hldpra2010}.
Virial expansion for balanced Fermi gases with equal spin populations
has already shown to be very useful, in the studies of both static
\cite{Ho,plbHorowitzQVE,lhdprl2009,lhdpra2010,lhdprb2010} and dynamic
properties \cite{hldprl2010,hldpra2010}. The third virial coefficient
of a unitary Fermi gas with infinitely large scattering length has
been calculated \cite{lhdprl2009} and consequently confirmed experimentally
\cite{ensNature}. The third virial coefficient of strongly interacting
Bose gases has been determined \cite{BedaqueB3}. The virial expansion
of single-particle spectral function \cite{hldprl2010} or dynamic
structure factor \cite{hldpra2010} has also been developed to understand,
respectively, the relevant measurements of momentum-resolved rf spectroscopy
\cite{jilaNature,jilaNP} or Bragg spectroscopy \cite{Veeravalli}.

Our generalization of quantum virial expansion to the imbalanced systems
gives an approximate means to study interesting phenomena associated
with the spin degree of freedom. For instance, for a strongly repulsively
interacting Fermi gas, itinerant ferromagnetic transition is expected
at low temperatures \cite{Stoner,Duine,LeBlanc,mitScience,Cui,Dong}.
However, the critical transition temperature is yet to be determined.
The use of virial expansion may allow us to estimate an upper bound
for the critical temperature. By expanding to the third order, we
find that there is no itinerant ferromagnetism as far as the quantum
virial expansion is applicable. Therefore, we show that for a strongly
repulsive Fermi gas in harmonic traps, the critical temperature should
be smaller than the Fermi temperature at the center of the trap. We
note that, virial expansion method has been used by Horowitz and Schwenk
to study the low density, spin-imbalanced neutron matter \cite{plbHorowitz,npaHorowitz}.
Their expansion was taken up to the second order, where the second
virial coefficient was calculated from the nucleon-nucleon scattering
phase shifts.

Our paper is organized as follows. In the next section, we show how
to expand the thermodynamic potential (or pressure in case of a uniform
system) of an imbalanced Fermi system in terms of virial coefficients.
The $n$-th virial coefficients can be calculated from the partition
functions of a cluster that contains up to $n$-particles. Different
spin configurations in a $n$-particle cluster give rise to different
virial coefficients. In Sec. III, we focus on the expansion to the
third order and calculate the finite temperature free energy of an
imbalanced Fermi gas as a function of spin imbalance. The possibility
of itinerant ferromagnetism in a strongly repulsively interacting
system is discussed. Sec. V is devoted to conclusions and some final
remarks.

\section{Quantum virial expansion for an imbalanced Fermi gas}

The starting point of quantum virial expansion is that at large temperatures
the chemical potential $\mu$ diverges to $-\infty$. Therefore, the
fugacity $z\equiv\exp(\mu/k_{B}T)\equiv\exp(\beta\mu)\ll1$ is a well-defined
small expansion parameter. We may expand the high-temperature thermodynamic
potential $\Omega$ of a quantum system in powers of fugacity, no
matter how strong the interaction strength would be.

For a two-component Fermi gas with balanced spin populations, it was
shown that \cite{lhdprl2009,lhdpra2010}, \begin{equation}
\Omega=-k_{B}TQ_{1}\left[z+b_{2}z^{2}+\cdots+b_{n}z^{n}+\cdots\right],\end{equation}
 where $Q_{n}=Tr_{n}[\exp(-{\cal H}/k_{B}T)]$ is the partition function
of a cluster containing $n$ particles and $b_{n}$ is the $n$-th
virial expansion coefficient. The trace $Tr_{n}$ takes into account
all the spin-configurations of $n$-particles and traces over all
the states of a proper symmetry. The virial coefficient $b_{n}$ was
found to take the form \cite{lhdprl2009,lhdpra2010}, \[
b_{2}=Q_{2}/Q_{1}-Q_{1}/2,\]
 and \begin{equation}
b_{3}=Q_{3}/Q_{1}-Q_{2}+Q_{1}^{2}/3,\quad etc.\end{equation}

In the presence of spin imbalance, it is necessary to introduce two
fugacities $z_{\uparrow}\equiv\exp(\mu_{\uparrow}/k_{B}T)$ and $z_{\downarrow}\equiv\exp(\mu_{\downarrow}/k_{B}T)$,
and to distinguish different spin-configurations. Quite generally,
we may write the thermodynamic potential as, \begin{equation}
\Omega=-k_{B}TQ_{1}\sum_{n=1}^{\infty}\sum_{k=0}^{n}z_{\uparrow}^{n-k}z_{\downarrow}^{k}b_{n,k},\end{equation}
 where $b_{n,k}$ is the $n$-th (imbalanced) virial coefficient contributed
by the configuration with $n-k$ spin-up fermions and $k$ spin-down
fermions. It is easy to see that the imbalanced virial coefficients
satisfy the relation $b_{n,k}=b_{n,n-k}$ and $\sum_{k=0}^{n}b_{n,k}=b_{n}$.

The calculation of $b_{n,k}$ is straightforward, following the standard
definition of thermodynamic potential. We rewrite the grand partition
function ${\cal Z}\equiv Tr\exp[-({\cal H}-\mu_{\uparrow}{\cal N}_{\uparrow}-\mu_{\downarrow}{\cal N}_{\downarrow})/k_{B}T]$
in the form, \begin{equation}
{\cal Z}=\sum_{n=0}^{\infty}\sum_{k=0}^{n}z_{\uparrow}^{n-k}z_{\downarrow}^{k}Q_{n,k},\end{equation}
 where $Q_{n,k}$ is the partition function of a cluster that contains
$n-k$ spin-up fermions and $k$ spin-down fermions. It is apparent
that due to the symmetry in spin configurations we have $Q_{n,k}=Q_{n,n-k}$.
The imbalanced cluster partition functions satisfy as well a sum rule
$\sum_{k=0}^{n}Q_{n,k}=Q_{n}$. By expanding the thermodynamic potential
$\Omega=-k_{B}T\ln{\cal Z}$ into powers of the two fugacities, the
imbalanced virial coefficients can then be expressed in terms of the
cluster partition function $Q_{n,k}$.

\subsection{Virial expansion up to the third order}

To be concrete, let us consider the imbalanced virial expansion up
to the third order. To this order, we may write the grand partition
function as ${\cal Z}=1+x_{1}+x_{2}+x_{3}$, where \[
x_{1}=z_{\uparrow}Q_{1,0}+z_{\downarrow}Q_{1,1},\]
 \[
x_{2}=z_{\uparrow}^{2}Q_{2,0}+z_{\uparrow}z_{\downarrow}Q_{2,1}+z_{\downarrow}^{2}Q_{2,2},\]
 and \begin{equation}
x_{3}=z_{\uparrow}^{3}Q_{3,0}+z_{\uparrow}^{2}z_{\downarrow}Q_{3,1}+z_{\uparrow}z_{\downarrow}^{2}Q_{3,2}+z_{\downarrow}^{3}Q_{3,3}.\end{equation}
 By introducing a symmetric cluster partition function $Q_{n}^{s}\equiv Q_{n,0}=Q_{n,n}$
and using the properties of $Q_{n,k}$, it is easy to show that $Q_{1}^{s}=Q_{1}/2$,
$Q_{2,1}=Q_{2}-2Q_{2}^{s}$, and $Q_{3,1}=Q_{3,2}=Q_{3}/2-Q_{3}^{s}$.
Using $\ln(1+x_{1}+x_{2}+x_{3})\simeq(x_{1}+x_{2}+x_{3})-(x_{1}^{2}+2x_{1}x_{2})/2+x_{1}^{3}/3$,
after some algebra we obtain $b_{n,k}$ ($k\leq n/2$), \begin{equation}
b_{1,0}=1/2,\end{equation}
 \begin{equation}
b_{2,0}=Q_{2}^{s}/Q_{1}-Q_{1}/8,\end{equation}
 \begin{equation}
b_{2,1}=Q_{2}/Q_{1}-2Q_{2}^{s}/Q_{1}-Q_{1}/4,\end{equation}
 \begin{equation}
b_{3,0}=Q_{3}^{s}/Q_{1}-Q_{2}^{s}/2+Q_{1}^{2}/24,\end{equation}
 and \begin{equation}
b_{3,1}=Q_{3}/(2Q_{1})-Q_{3}^{s}/Q_{1}-Q_{2}/2+Q_{2}^{s}/2+Q_{1}^{2}/8.\end{equation}
 The virial coefficients with $k\geq n/2$ can be obtained directly
since $b_{n,k}=b_{n,n-k}$.

In practice, it is convenient to consider the interaction effect on
the virial coefficients or the differences such as $\Delta Q_{n}=Q_{n}-Q_{n}^{(1)}$,
$\Delta b_{n}=b_{n}-b_{n}^{(1)}$, and $\Delta b_{n,k}=b_{n,k}-b_{n,k}^{(1)}$.
Here, the superscript {}``$1$'' denotes an ideal, non-interacting
system with the same fugacities and the operator {}``$\Delta$''
removes the non-interacting contribution. It is clear that the symmetric
cluster partition function $Q_{n}^{s}$ is not affected by interactions
since the interatomic interaction occurs only between fermions with
unlike spins. Thus, we have $\Delta b_{2,0}=\Delta b_{3,0}=0$, $\Delta b_{2,1}=\Delta(b_{2}-2b_{2,0})=\Delta b_{2}$
and $\Delta b_{3,1}=\Delta(b_{3}/2-b_{2,0})=\Delta b_{3}/2$. Accordingly,
we may rewrite the thermodynamic potential into the form (up to the
third order), \begin{equation}
\Omega=\Omega^{(1)}-k_{B}TQ_{1}\left[z_{\uparrow}z_{\downarrow}\Delta b_{2}+\frac{z_{\uparrow}^{2}z_{\downarrow}+z_{\uparrow}z_{\downarrow}^{2}}{2}\Delta b_{3}\right],\label{omega}\end{equation}
 where $\Omega^{(1)}$ is the thermodynamic potential of a non-interacting
Fermi gas.

\subsection{Thermodynamic potential $\Omega^{(1)}$}

The thermodynamic potential of an ideal, imbalanced Fermi gas is simply
the sum of thermodynamic potential of each component, $\Omega^{(1)}=\Omega^{(1)}(\mu_{\uparrow})+\Omega^{(1)}(\mu_{\downarrow})$.
In homogeneous space, where $Q_{1,H}=2V/\lambda^{3}$, the single-component
thermodynamic potential $\Omega_{H}^{(1)}(\mu_{\sigma})$ takes the
form ($\sigma=\uparrow,\downarrow$), \begin{equation}
\Omega_{H}^{(1)}=-V\frac{k_{B}T}{\lambda^{3}}\frac{2}{\sqrt{\pi}}\int\limits _{0}^{\infty}t^{1/2}\ln\left(1+z_{\sigma}e^{-t}\right)dt,\end{equation}
 where $V$ is the volume and $\lambda\equiv[2\pi\hbar^{2}/(mk_{B}T)]^{1/2}$
is the thermal wavelength. We have used the subscript {}``$H$''
to denote the result in the homogeneous space.

In the presence of a harmonic trap $V_{T}(r)=m\omega^{2}r^{2}/2$,
we consider the thermodynamic limit with a large number of fermions.
The non-interacting thermodynamic potential $\Omega_{T}^{(1)}(\mu_{\sigma})$
is then given semiclassically by ($\beta\equiv1/k_{B}T$), \begin{eqnarray}
\Omega_{T}^{(1)} & = & -\frac{1}{\beta}\int\frac{d{\bf r}d{\bf k}}{\left(2\pi\right)^{3}}\ln\left[1+e^{-\beta\left(\frac{\hbar^{2}k^{2}}{2m}+\frac{m}{2}\omega^{2}r^{2}-\mu_{\sigma}\right)}\right],\\
 & = & -\frac{\left(k_{B}T\right)^{4}}{\left(\hbar\omega\right)^{3}}\frac{1}{2}\int\limits _{0}^{\infty}t^{2}\ln\left(1+z_{\sigma}e^{-t}\right)dt.\end{eqnarray}
 Accordingly, we have $Q_{1,T}=2\left(k_{B}T\right)^{3}/\left(\hbar\omega\right)^{3}$.
Here, the subscript {}``$T$'' denotes the result in a harmonic
trap.

\subsection{Virial coefficients $\Delta b_{2}$ and $\Delta b_{3}$}

The calculation of virial coefficients requires the knowledge of full
energy spectrum of a few-body system. By utilizing the exact two-
and three-fermion solutions, we have calculated the second and third
virial coefficients for a strongly attractively or repulsively interacting
Fermi gas \cite{lhdpra2010}. Here, the terminology of {}``a strongly
attractively interacting Fermi gas'' refers to a resonant Fermi gas
near a Feshbach resonance with the inclusion of the molecular states,
while the terminology of {}``a strongly repulsively inteacting Fermi
gas'' means a resonant Fermi gas with the molecular branch excluded.
In other words, all fermions in the strongly \emph{repulsively} interacting
Fermi gas populate only on the upper, atomic branch of the energy
levels near a Feshbach resonance. At unitarity, three resonantly interacting
fermions in an isotropic harmonic trap can be solved exactly using
hyperspherical coordinates \cite{Werner1,Werner2}, with the relative
energy given by $E_{rel}=(2q+s_{l,n}+1)\hbar\omega$, where $q$ and
$s_{l,n}$ ($l$ the relative angular momentum and $n$ the number
of nodes) are respectively the good quantum numbers in the hyperradius
and hyperangle equations. One can show that the molecular branch corresponds
to the states with $s_{l,n=0}$ \cite{lhdpra2010}. These states should
be excluded for the {}``repulsive'' case. We refer to the Sec. IIIB
of Ref. \cite{lhdpra2010} for a detailed discussion about this. 

In the following, we shall focus on the unitarity limit with infinitely
large scattering length, where the virial coefficients are universal
and temperature independent. The coefficients for a homogeneous gas
are given by \cite{lhdpra2010}, \begin{eqnarray}
\Delta b_{2,H}^{att} & = & +1/\sqrt{2},\\
\Delta b_{2,H}^{rep} & = & -1/\sqrt{2},\end{eqnarray}
 and \begin{eqnarray}
\Delta b_{3,H}^{att} & \simeq & -0.35501298,\\
\Delta b_{3,H}^{rep} & \simeq & +1.8174.\end{eqnarray}
 Here, the superscript {}``att'' (or {}``rep'') means the coefficient
of an attractively (or repulsively) interacting Fermi gas. The third
virial coefficient of an attractive unitary Fermi gas, $\Delta b_{3,H}^{att}\simeq-0.355$,
was recently confirmed experimentally at École Normale Supérieure
(ENS), Paris \cite{ensNature}.

For a trapped Fermi gas, we instead have, \begin{eqnarray}
\Delta b_{2,T}^{att} & = & +1/4,\\
\Delta b_{2,T}^{rep} & = & -1/4,\end{eqnarray}
 and \begin{eqnarray}
\Delta b_{3,T}^{att} & \simeq & -0.06833960,\\
\Delta b_{3,T}^{rep} & \simeq & 0.34976.\end{eqnarray}

We note that, in the unitarity limit the homogeneous and trapped virial
coefficients are connected universally by the relation \cite{lhdprl2009,lhdpra2010},
\begin{equation}
b_{n,T}=\frac{b_{n,H}}{n^{3/2}}.\label{BnTrapVsBnHomo}\end{equation}
The factor of $n^{-3/2}$ reduction in harmonic traps is simply due
to the higher density of states in traps. To see this, let us consider
the thermodynamic potential of a harmonic trapped Fermi gas in the
local density approximation $\Omega=\int d{\bf r}\Omega({\bf r})$,
where $\Omega({\bf r})$ is the local thermodynamic potential \begin{equation}
\Omega({\bf r})\propto z\left({\bf r}\right)+b_{2,H}z^{2}\left({\bf r}\right)+b_{3,H}z^{3}\left({\bf r}\right)+\cdots.\end{equation}
 Here, the local fugacity $z\left({\bf r}\right)=z\exp[-V({\bf r})/k_{B}T]$
is determined by the local chemical potential $\mu({\bf r})=\mu-V({\bf r})$.
The spatial integration immediately leads to Eq. (\ref{BnTrapVsBnHomo}).
It is obivious that the suppressed virial coefficients in harmonic
traps imply a weaker interaction effect and consequently a better
convergence for virial expansion.

\subsection{Virial expansion of high-$T$ free energy}

We are now ready to study the high temperature thermodynamics of an
imbalanced, strongly interacting Fermi gas. With the virial expansion
of thermodynamic potential Eq. (\ref{omega}), we solve the standard
thermodynamic relations $N_{\uparrow}=-\partial\Omega/\partial\mu_{\uparrow}$
and $N_{\downarrow}=-\partial\Omega/\partial\mu_{\downarrow}$ for
the two fugacities $z_{\uparrow}$ and $z_{\downarrow}$, at a given
reduced temperature $\tau=T/T_{F}$ and a given spin imbalance $P=(N_{\uparrow}-N_{\downarrow})/N$.
Here, $T_{F}$ is the Fermi temperature. It is given by $T_{F}=\hbar^{2}(3\pi^{2}n)^{2/3}/(2m)/k_{B}$
in the homogeneous space and by $T_{F}=(3N)^{1/3}\hbar\omega/k_{B}$
in a harmonic trap.

In the homogeneous space, it is easy to show that we can define a
dimensionless number density $\tilde{n}=n\lambda^{3}=8/(3\sqrt{\pi}\tau^{3/2})$,
$\tilde{n}_{\uparrow}=(1+p)\tilde{n}/2$, and $\tilde{n}_{\downarrow}=(1-p)\tilde{n}/2$.
We then rewrite the number equations into dimensionless forms, \begin{eqnarray}
\tilde{n}_{\uparrow} & = & n\left(z_{\uparrow}\right)+z_{\uparrow}z_{\downarrow}2\Delta b_{2}+\left(2z_{\uparrow}^{2}z_{\downarrow}+z_{\uparrow}z_{\downarrow}^{2}\right)\Delta b_{3},\\
\tilde{n}_{\downarrow} & = & n\left(z_{\downarrow}\right)+z_{\uparrow}z_{\downarrow}2\Delta b_{2}+\left(z_{\uparrow}^{2}z_{\downarrow}+2z_{\uparrow}z_{\downarrow}^{2}\right)\Delta b_{3},\end{eqnarray}
 where $n\left(z\right)\equiv(2/\sqrt{\pi})\int\nolimits _{0}^{\infty}\sqrt{t}[ze^{-t}/\left(1+ze^{-t}\right)]dt$.
We can obtain the two fugacities by solving the coupled number equations.
Then, we calculate the free energy $F=\Omega+\mu_{\uparrow}N_{\uparrow}+\mu_{\downarrow}N_{\downarrow}$.
The free energy per particle in units of Fermi energy $E_{F}$ is
given by, \begin{eqnarray}
\frac{F}{NE_{F}} & = & \tau\left[\frac{1+p}{2}\ln z_{\uparrow}+\frac{1-p}{2}\ln z_{\downarrow}\right]-A\left[f\left(z_{\uparrow}\right)\right.\nonumber \\
 &  & \left.+f\left(z_{\downarrow}\right)+z_{\uparrow}z_{\downarrow}2\Delta b_{2}+\left(z_{\uparrow}^{2}z_{\downarrow}+z_{\uparrow}z_{\downarrow}^{2}\right)\Delta b_{3}\right],\end{eqnarray}
 where $f\left(z\right)\equiv(2/\sqrt{\pi})\int\nolimits _{0}^{\infty}\sqrt{t}\ln\left(1+ze^{-t}\right)dt$
and the constant $A=3\sqrt{\pi}\tau^{5/2}/8$.

In a harmonic trap, we have very similar equations with exactly the
same structure. However, the dimensionless number density is now given
by, $\tilde{n}=1/(3\tau^{3})$. The function $n(z)$ and $f(z)$ shall
take the form, \begin{eqnarray}
n\left(z\right) & \equiv & \frac{1}{2}\int\nolimits _{0}^{\infty}\frac{ze^{-t}}{\left(1+ze^{-t}\right)}t^{2}dt,\\
f\left(z\right) & \equiv & \frac{1}{2}\int\nolimits _{0}^{\infty}\ln\left(1+ze^{-t}\right)t^{2}dt,\end{eqnarray}
 respectively. Finally, the constant $A$ in free energy should be
read as $A=3\tau^{4}$.

\section{Results and discussions}

We discuss separately the high-temperature free energy of a unitary
Fermi gas in homogeneous space and in a harmonic trap.

\begin{figure}
\begin{centering}
\includegraphics[clip,width=0.4\textwidth]{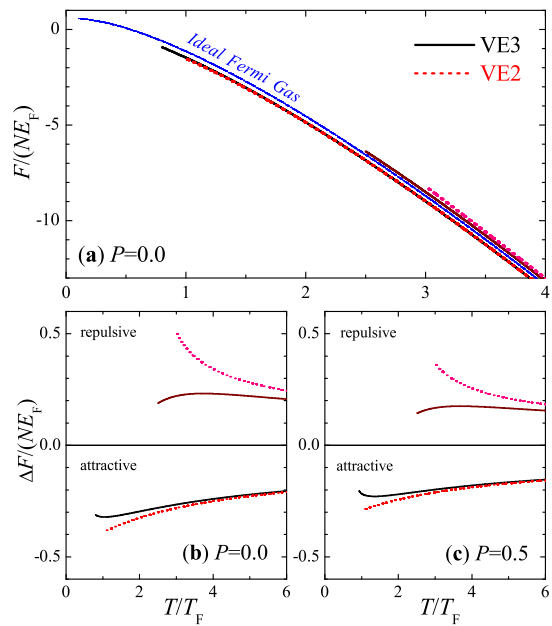} 
\par\end{centering}

\caption{(color online) Temperature dependence of the free energy of a homogeneous
unitary Fermi gas, calculated up to the second-order (dashed lines)
or third-order (solid lines) using the quantum virial expansion theory.
For comparison, we show the ideal, non-interacting free energy by
a thin dot-dashed line. As anticipated, the free energy of an attractive
or a repulsive Fermi gas is always smaller or larger than that of
an ideal Fermi gas, respectively.}

\label{fig1} 
\end{figure}

\subsection{Free energy of a homogeneous unitary Fermi gas}

Fig. 1 presents the virial expansion prediction for the free energy
as a function of temperature, for a balanced unitary Fermi gas or
for a spin imbalanced system. For comparison, we show also the free
energy of an ideal, non-interacting Fermi gas, $F^{(1)}$.To illustrate
clearly the interaction effect, we show in Figs. 1b and 1c the residue
free energy or the interaction energy $\Delta F=F-F^{(1)}$, by subtracting
the non-interacting background $F^{(1)}$. As shown in Fig. 1a and
in the lower part of Fig. 1b for a balanced Fermi gas, in the presence
of attractive interactions the second- and third-order virial expansion
predict quantitatively the same free energy down to the Fermi degenerate
temperature $T_{F}$. This suggests a very broad temperature window
for applying the quantum virial expansion method. In case of repulsive
interactions, however, the good agreement from the expansions at different
order is found only at $T>5T_{F}$, indicating that the virial expansion
for a repulsively interacting Fermi gas is severely restricted. The
similar applicability of virial expansion is observed as well for
an imbalanced Fermi gas with a general spin imbalance $P$. On the
other hand, by comparing the interaction energy at $P=0$ (Fig. 1b)
and $P=0.5$ (Fig. 1c), a finite spin imbalance reduces the interaction
energy as expected, although the reduction is not so significant.

\begin{figure}
\begin{centering}
\includegraphics[clip,width=0.4\textwidth]{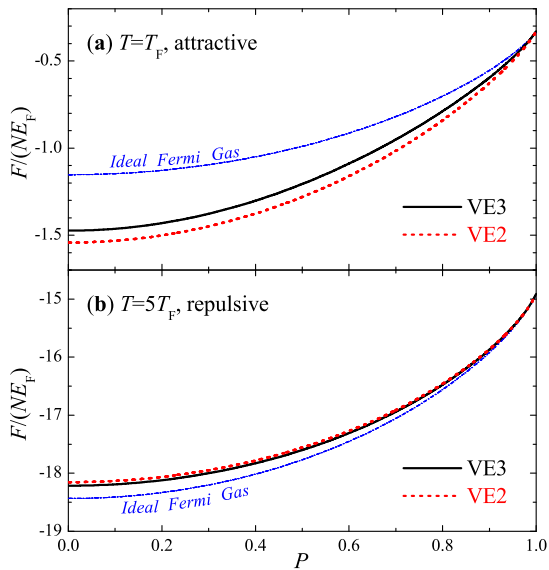} 
\par\end{centering}

\caption{(color online) Free energy of a homogeneous unitary Fermi gas as a
function of spin imbalance at a given temperature (as indicated),
for attractive interactions or for repulsive interactions. The exact
free energy could lie between the second- and third-order virial expansion
predictions.}

\label{fig2} 
\end{figure}

Fig. 2 reports the free energy as a function of the spin imbalance
at a given temperature, for an attractive unitary Fermi gas (upper
panel, Fig. 2a) or for a repulsive unitary Fermi gas (lower panel,
Fig. 2b). The exact free energy would be bounded by the two successive
expansion predictions and therefore could lie in between. With increasing
the spin imbalance, the free energy of interacting Fermi gases approaches
to the ideal, fully polarized limit. The reduction of interaction
energy is clearly visible. For the case with repulsive interactions
at $T=5T_{F}$, where we believe the quantitative applicability of
quantum virial expansion, the minimum of the free energy occurs at
the balanced limit of $P=0$. This implies that there is no itinerant
ferromagnetism at such a large temperature $5T_{F}$. It therefore
gives an upper bound for the ferromagnetic transition temperature
of a homogeneous Fermi gas at unitarity.

\begin{figure}
\begin{centering}
\includegraphics[clip,width=0.4\textwidth]{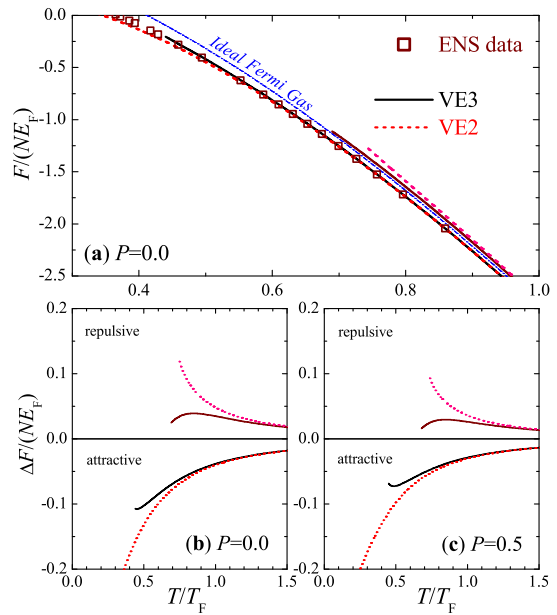} 
\par\end{centering}

\caption{(color online) Temperature dependence of the free energy of a trapped
unitary Fermi gas. We use the same notations as in Fig. 1. The experimental
data for an attractive trapped Fermi gas at unitarity, reported by
Salomon group at ENS \cite{ensNature}, are shown by squares.}

\label{fig3} 
\end{figure}

\begin{figure}
\begin{centering}
\includegraphics[clip,width=0.4\textwidth]{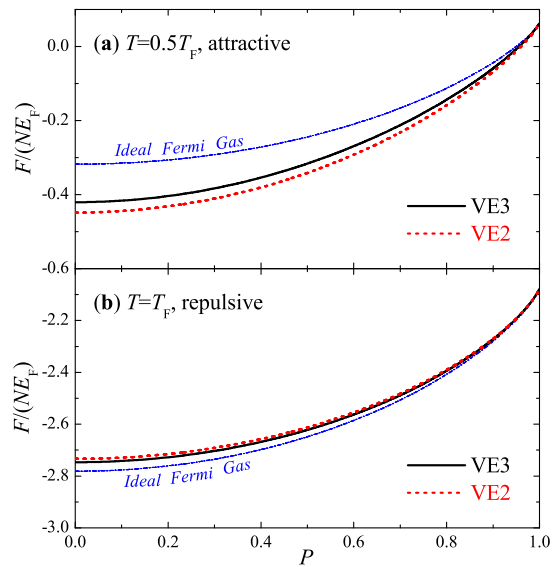} 
\par\end{centering}

\caption{(color online) Free energy of a trapped unitary Fermi gas as a function
of spin imbalance at a given temperature as indicated. The same notations
as in Fig. 2 have been used.}

\label{fig4} 
\end{figure}

\subsection{Free energy of a trapped unitary Fermi gas}

In analogy with the uniform free energy in Figs. 1 and 2, we show
in Figs. 3 and 4, respectively, the dependence of the free energy
on temperature and on spin imbalance. Due to the existence of harmonic
traps and hence a factor of $n^{-3/2}$ reduction for the $n$-th
virial coefficient as shown in Eq. (\ref{BnTrapVsBnHomo}), the applicability
of quantum virial expansion is greatly extended. At the third order,
we estimate that the virial expansion is now quantitatively reliable
down to $0.5T_{F}$ for an attractive Fermi gas and down to $T_{F}$
for a repulsive Fermi gas.

We estimate also the upper bound for the ferromagnetic transition
temperature of a trapped repulsively interacting Fermi gas at unitarity.
As seen from Fig. 4b, the trapped free energy does not exhibit any
signal for itinerant ferromagnetism at the Fermi degenerate temperature
$T_{F}$. We thus conclude that $(T_{c})_{upper}<T_{F}$ for a trapped
unitary Fermi gas.

\subsection{Absence of itinerant ferromagnetism in the third order virial expansion}

The absence of itinerant ferromagnetism in our virial expansion may
be analytically understood. To the third order, we calculate the two
by two susceptibility matrix ${\cal S}=(\partial n_{\sigma}/\partial\mu_{\sigma^{\prime}})$.
A divergent or negative spin susceptibility $\chi_{S}=(\partial\delta n/\partial\delta\mu)$
in the balanced limit of $P=0$ would signal the instability to the
formation of ferromagneitc order. The calculation of the susceptibility
matrix and $\chi_{S}$ is straightforward. We find that for a homogeneous
Fermi gas at unitarity, \begin{equation}
{\cal S}\left(P=0\right)=\frac{1}{k_{B}T\lambda^{3}}\left[\begin{array}{ll}
A & B\\
B & A\end{array}\right],\end{equation}
 where \begin{equation}
A=\frac{2}{\sqrt{\pi}}\int\limits _{0}^{\infty}\frac{\sqrt{t}ze^{-t}}{\left(1+ze^{-t}\right)^{2}}dt+2z^{2}\Delta b_{2,H}+5z^{3}\Delta b_{3,H}\end{equation}
 and \begin{equation}
B=2z^{2}\Delta b_{2,H}+4z^{3}\Delta b_{3,H}.\end{equation}
The spin susceptibility $\chi_{S}=2(A-B)/(k_{B}T\lambda^{3})$ is
then given by,\begin{equation}
\chi_{S}=\frac{2}{k_{B}T\lambda^{3}}\left[\frac{2}{\sqrt{\pi}}\int\limits _{0}^{\infty}\frac{\sqrt{t}ze^{-t}}{\left(1+ze^{-t}\right)^{2}}dt+z^{3}\Delta b_{3,H}\right].\label{spinkappa}\end{equation}
It is easy to see that, because of a positive third virial coefficient
$\Delta b_{3,H}^{rep}$, the spin susceptibility of a repulsive unitary
Fermi gas would always be finite and positive, for a finite fugacity.
Therefore, there must be no ferromagnetic transition if we restrict
the quantum virial expansion up to the third order. The same observation
is found for a trapped repulsive Fermi gas at unitarity.

\begin{figure}
\begin{centering}
\includegraphics[clip,width=0.4\textwidth]{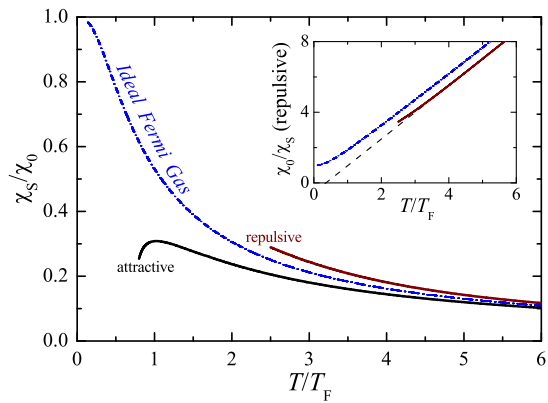} 
\par\end{centering}

\caption{(color online) Spin susceptibility of a homogeneous Fermi gas at unitarity,
normalized by the $T=0$ ideal Fermi gas susceptibility value $\chi_{0}=3n/(2\epsilon_{F})$,
with $n$ the density and $\epsilon_{F}$ the Fermi energy. The spin
susceptibility of a repulsive or attractive Fermi gas at unitarity
lies above or below the ideal Fermi Gas result (dot-dashed line).
The inset shows the inverse of the spin susceptibility of a repulsive
unitary Fermi gas. A linear extrapolation of the high-$T$ virial
expansion result at $T/T_{F}=4\sim6$ to low temperatures indicates
a ferromagnetic transition temperature of $T_{c}\sim0.4T_{F}$.}

\label{fig5} 
\end{figure}

\subsection{Third order expansion of the spin susceptibility for a homogeneous
Fermi gas at unitarity}

Figure 5 reports the numerical result of Eq. (\ref{spinkappa}) for
a homogeneous repulsive and attractive Fermi gas at unitarity, where
the fugacity $z$ is solved consistently to the third order expansion
in the number equation.

The spin susceptibility of a repulsive unitary Fermi gas is larger
than that of an ideal, non-interacting Fermi gas, as expected. Accordingly,
its inverse $\chi_{0}/\chi_{S}$ is smaller (see the inset). This
might be a high-temperature indication of the ferromagetic instability.
The transition temperature is conveniently determined from the condition
$\chi_{S}^{-1}(T_{c})=0$. By linearly extraplating the high-temperature
result of $\chi_{S}^{-1}$ (in the window of $T=4T_{F}\sim6T_{F}$)
to low temperatures, we estimate very roughly that $T_{c}\sim0.4T_{F}$.

The spin susceptibility of an attractive unitary Fermi gas shown in
Fig. 5 is also of interest. At finite temperatures, it is related
to the measurement of the thermal spin flucutations: \begin{equation}
\frac{\triangle\left(N_{\uparrow}-N_{\downarrow}\right)^{2}}{N}=k_{B}T\frac{\chi_{S}}{n}.\label{fluct}\end{equation}
A shot noise measurement of the spin fluctuations therefore could
be used as a sensitive thermometry for strongly interacting Fermi
gases \cite{Zhou}, provided that the spin susceptibility is known.
This seems to be now accessible, since the shot noise measurements
of the density fluctuations in a weakly interacting Fermi gas have
already been demonstrated very recently \cite{Mueller,Sanner}. We
find that the spin susceptibility of an attractive Fermi gas at unitarity
is strongly suppressed with respect to the ideal Fermi gas result,
even well above the degenerate temperature $T_{F}$. At $T=T_{F}$,
the reduction is about $40\%$.

\section{Conclusions and remarks}

In conclusions, we have presented a quantum virial expansion theory
for the thermodynamics of strongly interacting, spin-population imbalanced
Fermi gases. The (imbalanced) virial coefficients are calculable through
the cluster partition function with different spin configurations.
We have shown the virial expansion of thermodynamics up to the third
order and have discussed in detail the numerical results for the high-temperature
free energy of a homogeneous or trapped Fermi gas at unitarity. The
predicted free energy may confront future high-precision thermodynamic
measurements.

An interesting application of our virial expansion theory is the determination
of an upper bound for the ferromagnetic transition temperature of
a repulsive Fermi gas at unitarity. We have proven numerically and
analytically that, up to the third order quantum virial expansion
fails to predict itinerant ferromagnetism. As the third-order virial
expansion is quantitatively applicable at $T>T_{F}$ for a trapped
repulsively interacting Fermi gas at unitarity, we therefore have
estimated an upper bound $(T_{c})_{upper}=T_{F}$ for the critical
ferromagnetic temperature in harmonic traps. Further improvement of
the upper bound requires the calculation of higher-order (imbalanced)
virial coefficients, by using few-fermion solutions as the input \cite{Stecher,Blume,Daily}. 

Much wider applications of the quantum virial expansion method for
strongly interacting quantum gases seem to be feasible. In the near
future, we are interested in investigating the virial expansion of
the Tan's contact \cite{Tan,sutTanRelation}, the three-body recombination
rate \cite{Esry,Bedaque}, and the quantum viscosity \cite{Taylor,Enss}.
\begin{acknowledgments}
This work was supported in part by the ARC Centres of Excellence for
Quantum-Atom Optics (ACQAO), ARC Discovery Project No. DP0984522 and
No. DP0984637, NSFC Grant No. 10774190, and NFRPC Grant No. 2006CB921404
and No. 2006CB921306. \end{acknowledgments}

\end{document}